\documentclass[twocolumn,showpacs,preprintnumbers,amsmath,amssymb,prb,superscriptaddress]{revtex4-1}
\usepackage{epsfig}
\usepackage{graphicx}
\usepackage{dcolumn}
\usepackage{bm}
\usepackage{color} 

\begin{document}

\title{
Optical evidence of type-II Weyl semimetals MoTe$_2$ and WTe$_2$
}
\author{Shin-ichi Kimura}
\email{kimura@fbs.osaka-u.ac.jp}
\affiliation{Graduate School of Frontier Biosciences, Osaka University, Suita 565-0871, Japan}
\affiliation{Department of Physics, Graduate School of Science, Osaka University, Toyonaka 560-0043, Japan}
\author{Yuki Nakajima}
\affiliation{Department of Physics, Graduate School of Science, Osaka University, Toyonaka 560-0043, Japan}
\author{Zenjiro Mita}
\affiliation{Graduate School of Frontier Biosciences, Osaka University, Suita 565-0871, Japan}
\author{Rajveer Jha}
\affiliation{Department of Physics, Tokyo Metropolitan University, Hachioji, Tokyo 192-0397, Japan}
\author{Ryuji Higashinaka}
\affiliation{Department of Physics, Tokyo Metropolitan University, Hachioji, Tokyo 192-0397, Japan}
\author{Tatsuma D. Matsuda}
\affiliation{Department of Physics, Tokyo Metropolitan University, Hachioji, Tokyo 192-0397, Japan}
\author{Yuji Aoki}
\affiliation{Department of Physics, Tokyo Metropolitan University, Hachioji, Tokyo 192-0397, Japan}
\date{\today}
\begin{abstract}
The carrier dynamics and electronic structures of type-II Weyl semimetal candidates MoTe$_2$ and WTe$_2$ have been investigated by using temperature-dependent optical conductivity [$\sigma(\omega)$] spectra.
Two kinds of Drude peaks (narrow and broad) have been separately observed.
The width of the broad Drude peak increases with elevating temperature above the Debye temperature of about 130~K
in the same way as those of normal metals, on the other hand, the narrow Drude peak becomes visible below 80~K and the width is rapidly suppressed with decreasing temperature.
Because the temperature dependence of the narrow Drude peak is similar to that of a type-I Weyl semimetal TaAs, it was concluded to originate from Dirac carriers of Weyl bands.
The result suggests that the conductance has the contribution of two kinds of carriers, normal semimetallic and Dirac carriers, and this observation is an evidence of type-II Weyl semimetals of MoTe$_2$ and WTe$_2$.
The obtained carrier mass of the semimetallic bands and the interband transition spectra suggest the weak electron correlation effect in both materials.
\end{abstract}
%
\maketitle
%
\section{Introduction}
Materials with linear band dispersions at the Fermi level ($E_{\rm F}$) such as surface states of topological insulators and Dirac semimetals have been recently attracting attention because the very high mobility as well as high speed devices are expected thanks to the very low effective carrier mass of the linear band dispersion.
In the case of the space- or time-reversal symmetry breaking of Dirac semimetals, the spin degeneracy of the Dirac band is released in the momentum space, resulting in forming two Dirac bands with different spin directions at the symmetry points with respect to high symmetry axes.
These materials are named as Weyl semimetals.

Weyl semimetals are classed in two groups by the shape of the Fermi surface.
One named type-I exhibits point-like Fermi surfaces owing to Lorentz invariant band structures, and another named type-II has tilted Weyl cones appearing between electron and hole Fermi surfaces.
$1T'$-MoTe$_2$ and $1T'$-WTe$_2$ of transition metal dichalcogenides (TMD) are candidates of the type-II Weyl semimetals~\cite{Soluyanov2015,Sum2015}.
TMD forms four crystal structures, namely $2H$, $1T$, $1T'$, and $T_d$.
Theoretical predictions are that MoTe$_2$ is a semiconductor in $2H$ and $1T$ forms and the electronic structure has been revealed by optical measurements~\cite{Frindt1963,Grasso1972,Davey1972}, on the other hand, that in $1T'$ and $T_d$ forms is predicted as a semimetal~\cite{Gao2015}.

MoTe$_2$ with monoclinic $\beta$ structure ($P2_1m$, No.~11) at room temperature is regarded as a trivial metal, however, it has a phase transition at about 240~K with structure change to orthorhombic $1T'$ structure ($Pmn2_1$, No.~31).
The $1T'$-MoTe$_2$ (this is denoted as MoTe$_2$ hereafter) is believed to be a putative Weyl semimetal~\cite{Berger2018}.
MoTe$_2$ is attracting attention owing to the appearance of superconductivity at the temperatures of 0.1~K at ambient pressure and of 8.2~K at 11.7~GPa~\cite{Qi2016}.
Mo$_{1-x}$Nb$_x$Te$_2$ has anomalous enhancement of thermopower near the critical region between the polar and nonpolar metallic phases~\cite{Sakai2016}.
It is under debate whether these properties are related to the Weyl semimetallic band structure or not. 

WTe$_2$ also has orthorhombic crystal structure ($Pmn2_1$) below room temperature.
This material has a large positive magnetoresistance~\cite{Xiong2014},
and the electron and hole Fermi surfaces are the same size, which is consistent with other normal semimetals~\cite{Zhu2015}.
In addition to the quadratic magnetoresistance, a field-linear Nernst response distinguishes WTe$_2$ from other dilute metals~\cite{Zhu2015}.

Since both of MoTe$_2$ and WTe$_2$ have been theoretically predicted to be type-II Weyl semimetals~\cite{Soluyanov2015,ZWang2016},
the investigation of the relation of the electronic structures to interesting physical properties of these materials is important and has been performed so far by using angle-resolved photoemission spectroscopy (ARPES)~\cite{Jiang2017,Rhodes2017,Huang2016,Deng2016,Sakano2017,Tamai2016,Pletikosic2014,CWang2016,Sante2017,YWu2016,Wu2017,Bruno2016,Feng2016,Matsumoto2018}
 and quantum oscillation measurements~\cite{Yang2017,Berger2018,Luo2016,Ali2014,Li2017,Zhu2015}.
All measurements pointed out that electron and hole bands commonly exist on $E_{\rm F}$.
By surface-sensitive ARPES measurements, Fermi arc surface electronic states have been also observed.
The origin of the Fermi arcs has been concluded to be the existence of Weyl points in many experiments, but another opinion is also present~\cite{Bruno2016}, so the origin is still under debate.

The evidence of the existence of Weyl points can be observed in optical conductivity [$\sigma(\omega)$] spectra.
For instance, in the $\sigma(\omega)$ spectra of TaAs, which is one of type-I Weyl semimetals, the rapid decrease of the Drude width and the decrease of the Drude weight with decreasing temperature have been observed~\cite{Kimura2017,Xu2016}.
In normal semimetals, because carriers are mainly scattered by lattice vibrations, the electrical resistivity as well as the scattering rate $\gamma$ linearly increases with increasing temperature above the Debye temperature~\cite{AM}.
On the other hand, carriers at Weyl points have different temperature-dependent $\gamma$ from normal metals owing to the increase of mobility $\mu$ ($\gamma \propto 1/\mu$) at low temperatures~\cite{Shekhar2015,Kimura2017}.
Therefore, the temperature dependence of $\gamma$ is one of important indices to judge the existence of Dirac electrons of Weyl semimallic bands.

So far, temperature-dependent $\sigma(\omega)$ spectra of WTe$_2$ have been reported by Homes {\it et al.}~\cite{Homes2015}.
In the paper, owing to the model of existence of two kinds of carriers assumed as electrons and holes included in the Drude analysis, one of $\gamma$, which is equal to inverse relaxation time, becomes very small at low temperatures.
They have suggested that the carriers are related to the extremely large nonsaturating magnetoresistance, but they have not discussed the existence of Weyl electrons.
The anisotropic $\sigma(\omega)$ spectra of WTe$_2$ in the $ab$ plane have been reported, and the interband spectra can be well explained by {\it ab initio} band calculations~\cite{Frenzel2017}.
On the other hand, even though the optical spectra of $2H$-MoTe$_2$ have been reported previously~\cite{Frindt1963,Grasso1972,Davey1972},
the optical spectra of $1T'$-MoTe$_2$ have never been reported with our full knowledge.

In this paper, we report optical spectra of MoTe$_2$ and WTe$_2$, in which the evidence of Weyl electrons commonly appears in both materials.
As a result, the character of Weyl electrons can be identified by the analysis with the combination of two Drude curves, namely narrow and broad Drude components.
The $\gamma$ value of the narrow Drude component is strongly suppressed at low temperatures, which is similar to that of a type-I Weyl semimetal TaAs, suggesting the existence of Weyl electrons.
In addition, the electron correlation is discussed using the comparison of the experimentally obtained $\sigma(\omega)$ spectra with band calculations and calculated $\sigma(\omega)$ spectra.

\section{Experimental}
%
\begin{figure}[t]
\begin{center}
\includegraphics[width=0.45\textwidth]{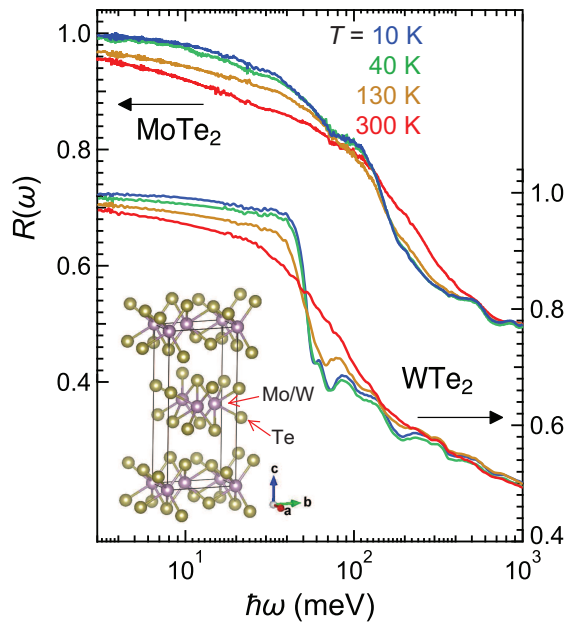}
\end{center}
\caption{
Temperature-dependent reflectivity [$R(\omega)$] spectra of MoTe$_2$ and WTe$_2$ in the $ab$ plane.
(Inset) Crystal structure of $1T'$-type MoTe$_2$ and WTe$_2$.
}
\label{fig:reflectivity}
\end{figure}

High-quality single-crystalline MoTe$_2$ and WTe$_2$ samples with the size of about $2\times0.5\times0.1$~mm$^3$ were synthesized by a self-flux method.
One of our samples, WTe$_2$, shows a large residual resistivity ratio of 1330 and very large magneto-resistance with obvious Shubnikov-de Haas oscillation~\cite{Jha2018,Onishi2018}. 
As-grown surfaces were measured for the optical reflectivity [$R(\omega)$] spectra.
Near-normal incident $R(\omega)$ spectra were acquired in a wide photon-energy range of 3~meV--30~eV to ensure accurate Kramers-Kronig analysis (KKA)~\cite{Kimura2013}.
Michelson-type and Martin-Puplett-type rapid-scan Fourier spectrometers were used at the photon energy $\hbar\omega$ regions of 8~meV--1.5~eV and 3--20~meV, respectively, with a feed-back positioning system to maintain the overall uncertainty level less than $\pm$0.5~\% in the temperature range of 10--300~K~\cite{Kimura2008}.
To obtain the absolute $R(\omega)$ values, {\it in-situ} evaporation method was adopted.
The obtained temperature-dependent $R(\omega)$ spectra of MoTe$_2$ and WTe$_2$ are shown in Fig.~\ref{fig:reflectivity}.
We could not recognize the structural phase transition of MoTe$_2$ at about 250~K by the $R(\omega)$ spectra.
This suggests that the change of the electronic structure is very small.
In the photon energy range of 1.2--30~eV, the $R(\omega)$ spectrum was only measured at room temperature by using synchrotron radiation, and connected to the spectra for $\hbar\omega \leq 1.5$~eV for the KKA.
All measurements were performed along the $ab$-plane.
In order to obtain $\sigma(\omega)$ spectra via KKA of $R(\omega)$ spectra, the $R(\omega)$ spectra were extrapolated below 3~meV with a Hagen-Rubens function, and above 30~eV with a free-electron approximation $R(\omega) \propto \omega^{-4}$~\cite{DG}.
The obtained $\sigma(\omega)$ spectra were compared with the spectra derived from LDA band structure calculations including spin-orbit coupling using the {\sc Wien2k} code~\cite{Wien2k}.
Lattice parameters of MoTe$_2$~\cite{ZWang2016} and WTe$_2$~\cite{Mar1992} with $1T'$ crystal structure were adopted.

\section{Results and Discussion}

\subsection{Comparison with band calculations}
\begin{figure}[t]
\begin{center}
\includegraphics[width=0.48\textwidth]{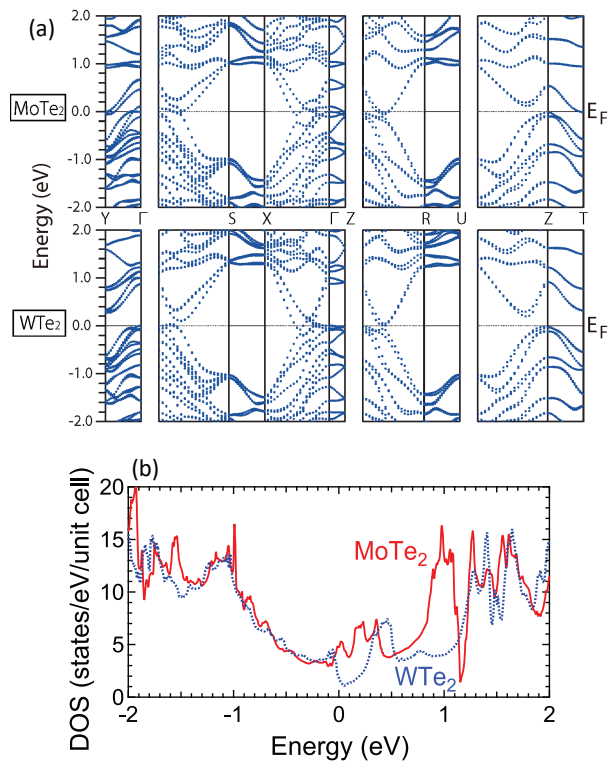}
\end{center}
\caption{
Band dispersions (a) and density of states (DOS, b) of MoTe$_2$ and WTe$_2$ in their orthorhombic $1T'$ structures ($Pmn2_1$).
}
\label{fig:band}
\end{figure}
Figure~\ref{fig:band} indicates the band dispersions and densities of states (DOS) of MoTe$_2$ and WTe$_2$.
Both materials are semimetals because electron and hole bands cross $E_{\rm F}$ at around the $\Gamma$ point.
This result is consistent with the previously reported band calculations~\cite{ZWang2016,Dawson1987}.
Weyl points are located near the symmetry line of $\Gamma-X$ (not shown in the figure).
In comparison between MoTe$_2$ and WTe$_2$, the dispersion curves are similar to each other, but both occupied and unoccupied states of MoTe$_2$ are closer to $E_{\rm F}$ than those of WTe$_2$, which can be also seen in DOS.
A peak at about 0.4~eV from $E_{\rm F}$ of WTe$_2$ shifts to about 0.2~eV in MoTe$_2$.
This causes an increase in DOS for MoTe$_2$ at around $E_{\rm F}$, i.e., the carrier density of MoTe$_2$ is expected to be higher than that of WTe$_2$.

\begin{figure}[t]
\begin{center}
\includegraphics[width=0.45\textwidth]{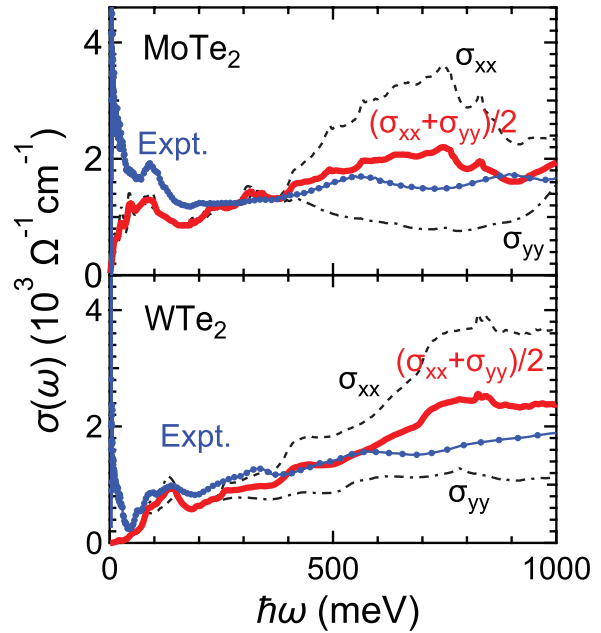}
\end{center}
\caption{
Experimental optical conductivity [$\sigma(\omega)$] spectra (marked lines) of MoTe$_2$ (a) and WTe$_2$ (b) compared with calculated $\sigma(\omega)$ spectra (bold, dashed, and dot-dashed lines) from the calculated band structures shown in Fig.~\ref{fig:band}.
Because of the experimental conditions of using non-polarized light, the experimental spectra should be compared with the averaged spectra [$(\sigma_{xx}+\sigma_{yy})/2$] of $\sigma_{xx}$ ($E \parallel a$) and $\sigma_{yy}$ ($E \parallel b$).
}
\label{fig:CalcOC}
\end{figure}

Calculated $\sigma(\omega)$ spectra obtained from band calculations in Fig.~\ref{fig:band}a in comparison with experimental ones are indicated in Fig.~\ref{fig:CalcOC}.
The experimental $\sigma(\omega)$ spectra in the $ab$-plane have been obtained using non-polarized light, so the spectra are compared with the average spectra [$(\sigma_{xx}+\sigma_{yy})/2$] of $\sigma_{xx}$ and $\sigma_{yy}$ along the $a$- and $b$-axes, respectively.
It should be noted that Drude structures appear in the experimental spectra in the photon energy region below 50~meV due to the existence of carriers, but it cannot be reproduced by the band calculations owing to no information on scattering rate.
In WTe$_2$, the experimental spectrum below the photon energy of 500~meV has been compared with the DFT-calculated spectrum by Homes {\it et al.}\cite{Homes2015}, and also with the DFT$+U$ calculation involving electron correlation ($U=2$~eV)~\cite{Sante2017}.
The latter provides better agreement with the experimental spectrum.

In the present data, even though the calculation without electron correlation has been performed, the spectrum below 500~meV, especially peaks at around 100~meV, seems to reproduce the experimental spectrum.
However, a peak at $\hbar\omega$ = 550~meV in MoTe$_2$ and two peaks at 300 and 600~meV in WTe$_2$ in the experimental spectra cannot be reproduced by our calculation.
There are peaks at 750~meV in MoTe$_2$ and at 400 and 800~meV in WTe$_2$ in the calculated spectra.
If these peaks in the calculated spectra are regarded to move to the experimental peaks by a self-energy effect owing to an electron correlation~\cite{Hewson1993},
the renormalization factor $z$ can be evaluated as $\sim0.73$ in MoTe$_2$ and $\sim0.75$ in WTe$_2$, which are almost consistent with each other.
These values are larger (this means smaller electron correlation) than those of normal heavy fermion compounds~\cite{Kimura2009},
but are similar to that of a valence fluctuation material YbAl$_2$ (the Kondo temperature $T_{\rm K}\sim2000$~K).
This is consistent with a relatively smaller correlation energy $U$ of 2.0~eV~\cite{Sante2017} than those of other strongly correlated electron systems~\cite{Imada1998}.
On the other hand, Frenzel {\it et al.} pointed out that the experimental $\sigma(\omega)$ spectra above the wavenumber of 5000 cm$^{-1}$ ($\sim0.6$~eV) in both $a$ and $b$ axes can be explained by the DFT calculations well~\cite{Frenzel2017}.
All of these results suggest the weak electron correlation intensity in MoTe$_2$ and WTe$_2$.

The overall intensity of the experimental $\sigma(\omega)$ spectrum of MoTe$_2$ is larger than that of WTe$_2$.
This implies that the DOS of MoTe$_2$ near $E_{\rm F}$ is higher than that of WTe$_2$.
This is consistent with the calculated DOS shown in Fig.~\ref{fig:band}b.
In addition, the experimental peaks at 100 and 50~meV in MoTe$_2$ can be reproduced in the calculation.
These peaks can be regarded to originate from flat structures in the bands near the $\Gamma$ point. 


\subsection{Drude analysis}
\begin{figure}[t]
\begin{center}
\includegraphics[width=0.45\textwidth]{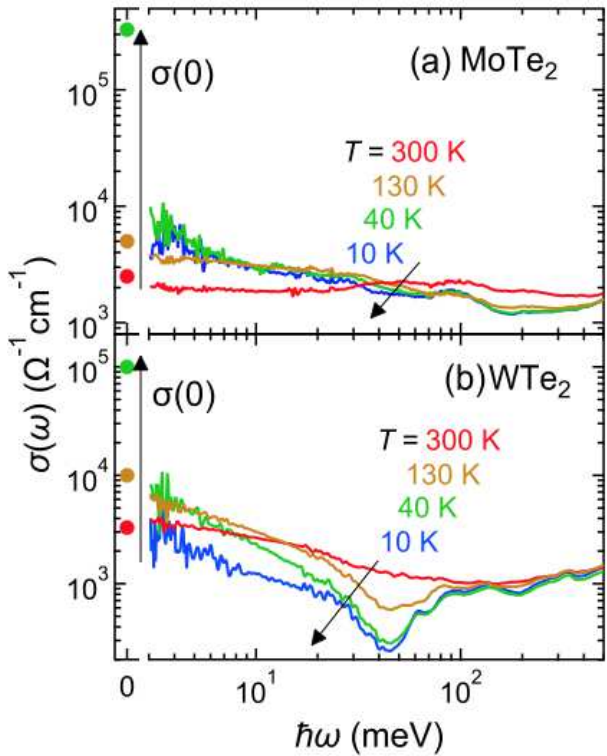}
\end{center}
\caption{
(a) Temperature-dependent optical conductivity [$\sigma(\omega)$] spectra of MoTe$_2$ (a) and WTe$_2$ (b) along the $ab$ plane in the photon energy ($\hbar\omega$) region below 500~meV and corresponding direct-current conductivity [$\sigma(0)$], which is consistent with $\sigma(\omega=0)$.
$\sigma(0)$ at 10~K is not shown because the value is out of range ($> 10^6$~$\Omega^{-1}$cm$^{-1}$)~\cite{Jha2018}.
}
\label{fig:Drude}
\end{figure}
Figure~\ref{fig:Drude} denotes $\sigma(\omega)$ spectra below the photon energy of 500~meV and the direct current [$\sigma(0)$] at the corresponding temperatures.
In both of MoTe$_2$ and WTe$_2$, $\sigma(0)$ values are located at the extrapolation to the lower energy side of the $\sigma(\omega)$ spectra  at temperatures above 130~K, however, $\sigma(0)$ values below 40~K become one or more orders higher than the extrapolation of the $\sigma(\omega)$ spectra.
It should be noted that $\sigma(0)$ at 10~K (higher than $10^6~\Omega^{-1}$cm$^{-1}$ in both materials~\cite{Jha2018}) is located out of range.
However, in our accessible lowest energy range above 3~meV, upturn structures in $\sigma(\omega)$ spectra with decreasing photon energy have been observed.
This suggests that the upturns are tails of the Drude curves connecting to the $\sigma(0)$ values, i.e., new conducting carriers emerge in addition to the conduction manifesting at high temperatures.

Considering these features, the $\sigma(\omega)$ spectra have been fitted by using the combination of two Drude functions as follows; 
\begin{equation}
\sigma(\omega)=\sum_{i=1}^2\dfrac{\sigma_{i}(0)}{\omega^2+\gamma_i^2},
\hspace{5mm}
\sigma_i{(0)} = \dfrac{N^*_i e^2\gamma_i}{m} \nonumber
\end{equation}
Here, $m$ denotes the electron rest mass, and $\sigma_{i}(0)$, $\gamma_i$, and $N^{*}_i$ are the direct current value, the damping constant of carriers, and the effective electron number of each component, respectively.

\begin{figure}[t]
\begin{center}
\includegraphics[width=0.45\textwidth]{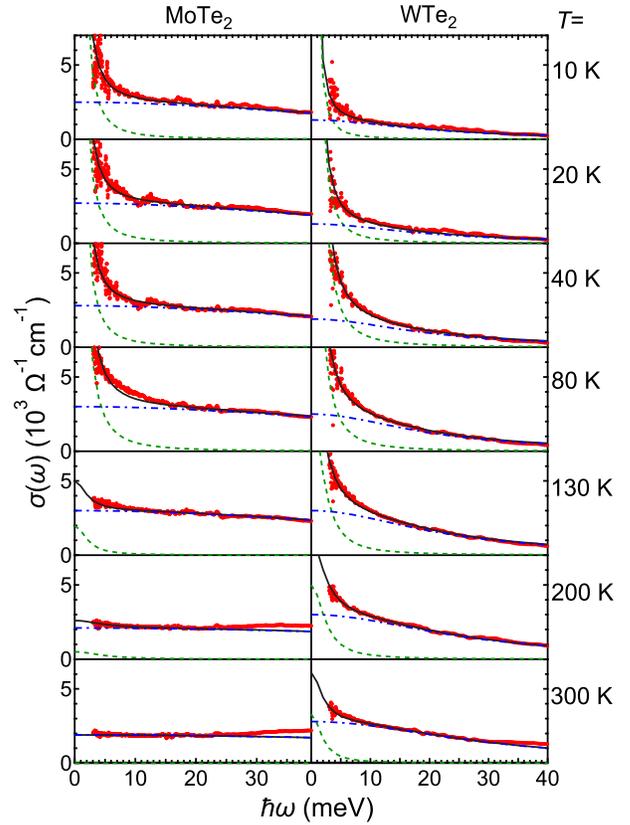}
\end{center}
\caption{
(Color online)
$\sigma(\omega)$ spectra of MoTe$_2$ and WTe$_2$ in the photon energy range below 40~meV at all measured temperatures (red marks).
The fitted Drude functions are denoted by dashed lines (D1) and dot-dashed lines (D2) and the sums are plotted by solid lines.
}
\label{fig:fitting}
\end{figure}
The fitting results are shown in Fig.~\ref{fig:fitting}.
In the fitting, the value of $\sigma_1(0)+\sigma_2(0)$ has been assumed to be equal to $\sigma(0)$.
In MoTe$_2$, at temperatures higher than 130~K, because the $\sigma(\omega)$ spectra are almost flat, a single broad Drude function (namely D2) can be applied.
Below 80~K, however, another Drude component (namely D1) with very narrow peak appears below the photon energy of 10~meV in addition to the D2 component.
D1 component becomes narrower with decreasing temperature.
In WTe$_2$, the temperature dependence is similar to that of MoTe$_2$, but narrow D1 component appears even at 130--300~K, which is the same result as shown by Homes {\it et al}.~\cite{Homes2015} 

\begin{figure}[t]
\begin{center}
\includegraphics[width=0.48\textwidth]{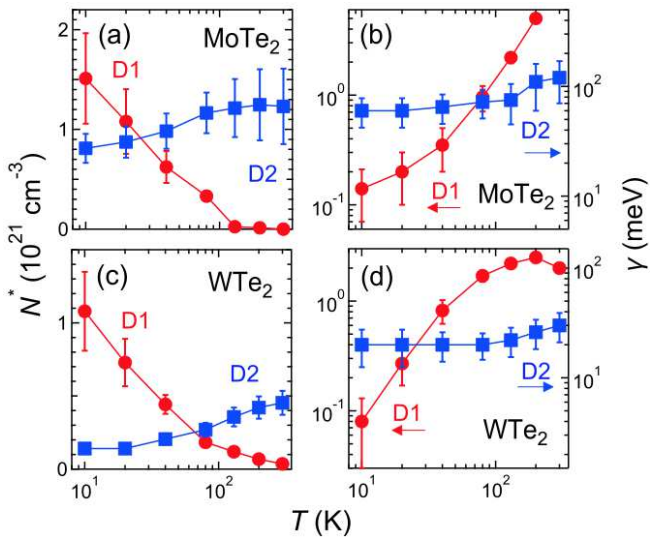}
\end{center}
\caption{
Drude fitting results in Fig.~\ref{fig:fitting} as a function of temperature.
(a) and (b) are the effective carrier density ($N^*$) and damping constant ($\gamma$), respectively, of D1 and D2 components of MoTe$_2$.
(c) and (d) are the same as (a) and (b), respectively, but of WTe$_2$.
}
\label{fig:fittingresults}
\end{figure}
Drude fitting results are shown in Fig.~\ref{fig:fittingresults}.
In the two Drude components, firstly, let us discuss the behavior of the broad Drude component D2. 
In MoTe$_2$, $N^*$ shown in Fig.~\ref{fig:fittingresults}a slightly decreases but is almost constant with decreasing temperature, 
but in WTe$_2$, shown in Fig.~\ref{fig:fittingresults}c, $N^*$ below 20~K becomes about three times smaller than that above 200~K.
The temperature dependence of $N^*$ originates from the shape of DOS near $E_{\rm F}$, i.e., 
WTe$_2$ has a large change in DOS near $E_{\rm F}$ in comparison with MoTe$_2$ as seen in Fig.~\ref{fig:band}b.
Therefore, the thermal excitation of electrons is considered to be the origin of the large temperature dependence of $\sigma(\omega)$ of WTe$_2$.
Even though the calculated temperature dependence of the carrier density using DOS in Fig.~\ref{fig:band}b has similar temperature dependence to the experimental $N^*$, however, the calculated carrier density increases only about 1.5 times with increasing temperature from 10 to 300~K (not shown here). 
This inconsistency suggests that the electron-hole asymmetry in the DOS should be more accentuated than the calculation shown in Fig.~\ref{fig:band}b.

The damping constant $\gamma$ of D2 in both materials is almost flat at temperatures below 80~K and slightly increases with increasing temperature above 130~K.
Since the $\gamma$ value originates from the scattering of carriers and the boundary temperature $T^*$ of about 100~K is consistent with the Debye temperature $\theta_D=133.8\pm0.6$~K of WTe$_2$~\cite{Callanan1992}, 
the temperature dependence of $\gamma$ of D2 originates from the scattering by phonons.
Below $T^*$, $\gamma$ becomes almost flat, which suggests the existence of constant scattering by impurities and/or defects.
This result is consistent with the behavior of other TMDs, i.e., the mobility ($\mu \propto  1/\gamma$) that has a tender slope below about 100~K changes to a rapid drop above the temperature~\cite{ZWu2016}.
Therefore, D2 component is concluded to originate from the semimetallic bands.
It should be noted that $N^*$ at 10~K evaluated as $1.5\times10^{20}$~cm$^{-3}$ is consistent with the sum of the electron and hole carriers obtained from the size of Fermi surfaces ($\sim1.35\times10^{20}$~cm$^{-3}$)~\cite{Zhu2015}.
This also suggests that D2 originates from semimetallic electron and hole bands and, in addition, the effective mass of carriers in the semimetallic bands is almost equal to the rest mass of electrons.

It should be noted that the carrier densities evaluated from the DOS of MoTe$_2$ and WTe$_2$ in Fig.~\ref{fig:band}b are $1.3\times10^{22}$~cm$^{-3}$ and $4.3\times10^{21}$~cm$^{-3}$, respectively, which are one order of magnitude larger than the experimentally obtained values.
According to the previous quantum oscillation and ARPES data, however, the experimental Fermi surface size is much smaller than the calculated one~\cite{Sante2017, Zhu2015, CWang2016, Pletikosic2014, YWu2016, Bruno2016}.
Therefore, the evaluated carrier density here is not inconsistent with other experiments.

Next, the narrow Drude component D1 is discussed.
In both materials, $N^*$ rapidly increases with decreasing temperature below about 100~K.
This change cannot be expected from the shape of DOS.
In addition, as pointed out by Homes {\it et al.}~\cite{Homes2015}, one of two semimetallic bands of the compensated metals is also a candidate.
However, since the sizes of the Fermi surfaces of electron and hole bands are the same as each other~\cite{Zhu2015} and the dispersions of the bands are also similar to each other~\cite{Pletikosic2014}, 
the Drude parameters of the electron and hole bands should be similar to each other.
Therefore, it is hard to believe that these two show different temperature dependences.
While, $\gamma$ of D1 rapidly decreases with decreasing temperature below 80~K in contrast to the flat of D2.
The temperature dependence of $\gamma$ is very similar to that of type-I Weyl semimetals, TaAs and TaP~\cite{Kimura2017}.
Therefore, D1 is considered to reflect the character of Dirac electrons at Weyl points of MoTe$_2$ and WTe$_2$.
From this point of view, the rapid increase of $N^*$ with decreasing temperature suggests that the rapid decrease of the effective mass $m^*$ because of $m^*/m=N_0/N^*$, where $N_0$ is the actual carrier density.
This suggests the presence of zero mass Dirac electrons.

The temperature dependence of the $\gamma$ of WTe$_2$ follows $\gamma \propto T^{1.6\pm0.05}$ below the temperature of 80~K as shown in Fig.~\ref{fig:fittingresults}b.
The extrapolation of the $\gamma$ to zero temperature is expected to be zero, in other words, the mobility becomes infinity at 0~K.
This suggests that the Weyl points are located just at $E_{\rm F}$.
It should be noted that the relation of $\gamma \propto T^{1.6\pm0.05}$ is close to the $T^2$-law of Weyl semimetals predicted by Hosur {\it et al.}~\cite{Hosur2012}.
On the other hand, the $\gamma$ of MoTe$_2$ has the temperature dependence of $\gamma \propto T^{1.5\pm0.05}$ at temperatures higher than 40~K, which is similar to that of WTe$_2$ below 80~K.
 The temperature dependence of $\gamma$ cannot be explained by the Fermi liquid theory but is similar to those of non-Fermi liquid materials~\cite{Degiorgi1999,Kimura2006}.
The temperature order of $\gamma$ of MoTe$_2$ decreases at temperatures lower than 40~K, which has the relation of $\gamma \propto T^{0.8\pm0.1}$.
This suggests that the Weyl points of MoTe$_2$ are located a little away (about 40~K $\sim$ 3~meV) from $E_{\rm F}$ because of a residual $\gamma$ value due to the existence of massive electrons and/or the electron correlation intensity increases with approaching $E_{\rm F}$.
An excitonic correlation might be effected to the origin~\cite{Matsumoto2018}.
The low-energy electron correlation near the Weyl points might effect to the physical properties such as superconductivity and high thermopower of MoTe$_2$.

The evidence for Dirac/Weyl semimetal phase in optical sepctra also appears in the power-law dependence of the interband optical conductivity$\sigma(\omega)$ spectra~\cite{Carbotte2016,Shao2019}.
The power-law dependence must appear in the energy region of Dirac linear bands.
In the case of MoTe$_2$ and WTe$_2$, it must appear below 50~meV with reference to band calculations.
However, since Drude component is dominant in the energy region, the power-law dependence may not be visible.
In other Weyl semimetals, TaAs and TaP, such power-law behaviors can be recognized because of weak Drude components.

\section{Conclusion}
To summarize, the electronic structures of type-II Weyl semimetal candidates MoTe$_2$ and WTe$_2$ have been investigated by optical conductivity $\sigma(\omega)$ spectra and LDA band calculations.
Obtained $\sigma(\omega)$ spectra can be explained by the band calculations with the renormalization factor of $0.74\pm0.01$.
Two kinds of Drude components have been observed below the photon energy of 40~meV.
One of the two components originates from the carriers of semimetallic electron and hole bands, 
the other from Dirac carriers at the Weyl points located at $E_{\rm F}$ that emerge only at low temperatures and the width of the Drude component become sharper continuously with decreasing temperature.
Therefore, MoTe$_2$ and WTe$_2$ have two kinds of carriers and are concluded to be type-II Weyl semimetals.

\section*{Acknowledgments}
Part of this work was performed under the Use-of-UVSOR Facility Program (BL7B, 2016--2017) of the Institute for Molecular Science.
This work was partly supported by JSPS KAKENHI (Grant No. 15H03676).


%

\begin{thebibliography}{99}
%
\bibitem{Soluyanov2015} 
A. A. Soluyanov, D. Gresch, Z. Wang, Q. S. Wu, M. Troyer, X. Dai, and B. A, Bernevig, 
Nature {\bf 527}, 495 (2015).
%
\bibitem{Sum2015} 
Y. Sun, S.-C. Wu, M. N. Ali, C. Felser, and B. Yan, 
Phys. Rev. B {\bf 92}, 161107 (2015).
%
\bibitem{Frindt1963}
R. F. Frindt,
J. Phys. Chem. Solids {\bf 24}, 1107 (1972).

\bibitem{Grasso1972}
V Grasso, G Mondio, and G Saitta,
J. Phys. C: Solid State Phys. {\bf 5}, 1101 (1972).

\bibitem{Davey1972} 
B. Davey and B. L. Evan,
phys. stat. sol. (a) {\bf 13}, 483 (1972).

\bibitem{Gao2015} 
G. Gao, Y. Jiao, F. Ma, Y. Jiao, E. Waclawik, and A. Du, 
J. Phys. Chem. {\bf 119}, 13124 (2015).

\bibitem{Berger2018} 
A. N. Berger, E. Andrade, A. Kerelsky, D. Edelberg, J. Li, Z. Wang, L. Zhang, J. Kim, N. Zaki, J. Avila, C. Chen, M. C. Asensio, S.-W. Cheong, B. A. Bernevig, and A. N. Pasupathy, 
npj Quantum Mater. {\bf 3}, 2 (2018).

\bibitem{Qi2016} 
Y. Qi, P. G. Naumov, M. N. Ali, C. R. Rajamathi, W. Schnelle, O. Barkalov, M. Hanfland, S.-C. Wu, C. Shekhar, Y. Sun, V. S\"u\ss, M. Schmidt, U. Schwarz, E. Pippel, P. Werner, R. Hillebrand, T F\"orster, E. Kampert, S. Parkin, R. J. Cava, C. Felser, B. Yan, and S. A. Medvedev, 
Nat. Commun. {\bf 7}, 11038 (2016).

\bibitem{Sakai2016} 
H. Sakai, K. Ikeura, M. S. Bahramy, N. Ogawa, D. Hashizume, J. Fujioka, Y. Tokura, and S. Ishiwata, 
Sci. Adv. {\bf 2}, e1601378 (2016).

\bibitem{Xiong2014} 
M. N. Ali, J. Xiong, S. Flynn, J. Tao, Q. D. Gibson, L. M. Schoop, T. Liang, N. Haldolaarachchige, M. Hirschberger, N. P. Ong, and R. J. Cava, 
Nature {\bf 514}, 205 (2014).

\bibitem{Zhu2015} 
Z. Zhu, X. Lin, J. Liu, B. Fauqu\'e, Q. Tao, C. Yang, Y. Shi, and K. Behnia, 
Phys. Rev. Lett. {\bf 114}, 176601 (2015).

\bibitem{ZWang2016} 
Z. Wang, D. Gresch, A. A. Soluyanov, W. Xie, S. Kushwaha, X. Dai, M. Troyer, R. J. Cava, and B. A. Bernevig, 
Phys. Rev. Lett. {\bf 117}, 056805 (2016).

\bibitem{Jiang2017} 
J. Jiang, Z. K. Liu, Y. Sun, H. F. Yang, C. R. Rajamathi, Y. P. Qi, L. X. Yang, C. Chen, H. Peng, C.-C. Hwang, S. Z. Sun, S.-K. Mo, I. Vobornik, J. Fujii, S. S. P. Parkin, C. Felser, B. H. Yan, and Y. L. Chen, 
Nat. Commun. {\bf 8}, 13973 (2017).

\bibitem{Rhodes2017} 
D. Rhodes, R. Sch\"onemann, N. Aryal, Q. Zhou, Q. R. Zhang, E. Kampert, Y.-C. Chiu, Y. Lai, Y. Shimura, G. T. McCandless, J. Y. Chan, D. W. Paley, J. Lee, A. D. Finke, J. P. C. Ruff, S. Das, E. Manousakis, and L. Balicas, 
Phys. Rev. B {\bf 96}, 165134 (2017).

\bibitem{Huang2016} 
L. Huang, T. M. McCormick, M. Ochi, Z. Zhao, M.-T. Suzuki, R. Arita, Y. Wu, D. Mou, H. Cao, J. Yan, N. Trivedi, and A. Kaminski,
Nat. Mater. {\bf 15}, 1155 (2016).

\bibitem{Deng2016}
K. Deng, G. Wan, P. Deng, K. Zhang, S. Ding, E. Wang, M. Yan, H. Huang, H. Zhang, Z. Xu, J. Denlinger, A. Fedorov, H. Yang, W. Duan, H. Yao, Y. Wu, S. Fan, H. Zhang, X. Chen, and S. Zhou,
Nat. Phys. {\bf 12}, 1105 (2016).

\bibitem{Sakano2017} 
M. Sakano, M. S. Bahramy, H. Tsuji, I. Araya, K. Ikeura, H. Sakai, S. Ishiwata, K. Yaji, K. Kuroda, A. Harasawa, S. Shin, and K. Ishizaka,
Phys. Rev. B {\bf 95}, 121101(R) (2017).

\bibitem{Tamai2016} 
A. Tamai, Q. S. Wu, I. Cucchi, F. Y. Bruno, S. Ricc\`o, T. K. Kim, M. Hoesch, C. Barreteau, E. Giannini, C. Besnard, A. A. Soluyanov, and F. Baumberger, 
Phys. Rev. X {\bf 6}, 031021 (2016).

\bibitem{Pletikosic2014} 
I. Pletikosi\'c, M. N. Ali, A. V. Fedorov, R. J. Cava, and T. Valla,
Phys. Rev. Lett. {\bf 113}, 216601 (2014).

\bibitem{CWang2016} 
C. Wang, Y. Zhang, J. Huang, S. Nie, G. Liu, A. Liang, Y. Zhang, B. Shen, J. Liu, C. Hu, Y. Ding, D. Liu, Y. Hu, S. He, L. Zhao, L. Yu, J. Hu, J. Wei, Z. Mao, Y. Shi, X. Jia, F. Zhang, S. Zhang, F. Yang, Z. Wang, Q. Peng, H. Weng, X. Dai, Z. Fang, Z. Xu, C. Chen, and X. J. Zhou,
Phys. Rev. B {\bf 94}, 241119(R) (2016).

\bibitem{Sante2017} 
D. Di Sante, P. K. Das, C. Bigi, Z. Erg\"onenc, N. G\"urtler, J. A. Krieger, T. Schmitt, M. N. Ali, G. Rossi, R. Thomale, C. Franchini, S. Picozzi, J. Fujii, V. N. Strocov, G. Sangiovanni, I. Vobornik, R. J. Cava, and G. Panaccione,
Phys. Rev. Lett. {\bf 119}, 026403 (2017).

\bibitem{YWu2016} 
Y. Wu, D. Mou, N. H. Jo, K. Sun, L. Huang, S. L. Bud’ko, P. C. Canfield, and A. Kaminski,
Phys. Rev. B {\bf 94}, 121113(R) (2016).

\bibitem{Wu2017} 
Y. Wu, N. H. Jo, D. Mou, L. Huang, S. L. Bud’ko, P. C. Canfield, and A. Kaminski, 
Phys. Rev. B {\bf 95}, 195138 (2017).

\bibitem{Bruno2016} 
F. Y. Bruno, A. Tamai, Q. S. Wu, I. Cucchi, C. Barreteau, A. de la Torre, S. McKeown Walker, S. Ricc\`o, Z. Wang, T. K. Kim, M. Hoesch, M. Shi, N. C. Plumb, E. Giannini, A. A. Soluyanov, and F. Baumberger,
Phys. Rev. B {\bf 94}, 121112(R) (2016).

\bibitem{Feng2016} 
B. Feng, Y.-H. Chan, Y. Feng, R.-Y. Liu, M.-Y. Chou, K. Kuroda, K. Yaji, A. Harasawa, P. Moras, A. Barinov, W. Malaeb, C. Bareille, T. Kondo, S. Shin, F. Komori, T.-C. Chiang, Y. Shi, and I. Matsuda, 
Phys. Rev. B {\bf 94}, 195134 (2016).

\bibitem{Matsumoto2018} 
R. Matsumoto, T. Sugimoto, T. Mizokawa, N. L. Saini, M. Arita, R. Jha, R. Higashinaka, T. D. Matsuda, and Y. Aoki,
Phys. Rev. B {\bf 98}, 205138 (2018).

\bibitem{Yang2017}
J. Yang, J. Colen, J. Liu, M. C. Nguyen, G.-w. Chern, and D. Louca,
Sci. Adv. {\bf 3}, eaao4949 (2017).

\bibitem{Luo2016}
X. Luo, F. C. Chen, J. L. Zhang, Q. L. Pei, G. T. Lin, W. J. Lu, Y. Y. Han, C. Y. Xi, W. H. Song, and Y. P. Sun,
Appl. Phys. Lett. {\bf 109}, 102601 (2016).

\bibitem{Ali2014}
M. N. Ali, J. Xiong, S. Flynn, J. Tao, Q. D. Gibson, L. M. Schoop, T. Liang, N. Haldolaarachchige, M. Hirschberger,N. P. Ong, and R. J. Cava,
Nature {\bf 514}, 205 (2014).

\bibitem{Li2017}
P. Li, Y. Wen, X. He, Q. Zhang, C. Xia, Z.-M. Yu, S. A. Yang, Z. Zhu, H. N. Alshareef, and X.-X. Zhang, 
Nat. Commun. {\bf 8}, 2150 (2017).

\bibitem{Kimura2017}
S. Kimura, H. Yokoyama, H. Watanabe, J. Sichelschmidt, V. S\"u\ss, M. Schmidt, and C. Felser,
Phys. Rev. B {\bf 96}, 075119 (2017).

\bibitem{Xu2016}
B. Xu, Y. M. Dai, L. X. Zhao, K. Wang, R. Yang, W. Zhang, J. Y. Liu, H. Xiao, G. F. Chen, A. J. Taylor, D. A. Yarotski, R. P. Prasankumar, and X. G. Qiu,
Phys. Rev. B {\bf 93}, 121110(R) (2016).

\bibitem{AM}
N. W. Aschcroft and N. D. Mermin,
{\it Solid State Physics},
(HRW International Editions, 1976), p. 523.

\bibitem{Shekhar2015}
C. Shekhar, A. K. Nayak, Y. Sun, M. Schmidt, M. Nicklas, I. Leermakers, U. Zeitler, Y. Skourski, J. Wosnitza, Z. Liu, Y. Chen, W. Schnelle, H. Borrmann, Y. Grin, C. Felser, and B. Yan.
Nat. Phys. {\bf 11}, 645 (2015).

\bibitem{Homes2015} 
C. C. Homes, M. N. Ali, and R. J. Cava,
Phys. Rev. B {\bf 92}, 161109(R) (2015).

\bibitem{Frenzel2017}
A. J. Frenzel, C. C. Homes, Q. D. Gibson, Y. M. Shao, K. W. Post, A. Charnukha, R. J. Cava, and D. N. Basov,
Phys. Rev. B {\bf 95}, 245140 (2017).

\bibitem{Jha2018}
R. Jha, R. Higashinaka, T. D. Matsuda, R. A. Ribeiro, and Yuji Aoki,
Physica B {\bf 536}, 68 (2018).

\bibitem{Onishi2018}
S. Onishi, R. Jha, A. Miyake, R. Higashinaka, T. D. Matsuda, M. Tokunaga, and Y. Aoki.
AIP Adv. {\bf 8}, 101330 (2018).

\bibitem{Kimura2013}
S. Kimura and H. Okamura, 
J. Phys. Soc. Jpn. {\bf 82}, 021004 (2013).
%
\bibitem{Kimura2008}
S. Kimura, JASCO Report {\bf 50}, 6 (2008). [in Japanese]
%
\bibitem{DG}
M. Dressel and G. Gr\"uner, {\it Electrodynamics of Solids} (Cambridge University Press, Cambridge, UK, 2002).
%
\bibitem{Wien2k}
P. Blaha, K. Schwarz, P. Sorantin, and S. B. Trickey, Comput. Phys. Commun. {\bf 59}, 399 (1990).
%
%
\bibitem{Mar1992}
A. Mar, S. Jobic, and J. A. Ibers,
J. Am. Chem. Soc. {\bf 114}, 8963 (1992).
%
\bibitem{Dawson1987}
W. G. Dawson and D. W. Bullet, J. Phys.: Solid State Phys. {\bf 20}, 6159 (1987).
%
\bibitem{Hewson1993}
A. C. Hewson, {\it The Kondo Problem to Heavy Fermions} (CambridgeUniversity Press, Cambridge, 1993).
%
\bibitem{Kimura2009}
S. Kimura, 
Phys. Rev. B {\bf80}, 073103 (2009).
%
\bibitem{Imada1998}
M. Imada, A. Fujimori, and Y. Tokura, Rev. Mod. Phys. {\bf 70}, 1039 (1998).
%
\bibitem{Callanan1992}
J. E. Callanan, G. A. Hope, R. D. Weir, and E. F. Westrum, Jr., 
J. Chem. Thermodynamics {\bf 24}, 627 (1992).
%
\bibitem{ZWu2016}
Z. Wu, S. Xu, H. Lu, A. Khamoshi, G.-B. Liu, T. Han, Y. Wu, J. Lin, G. Long, Y. He, Y. Cai, Y. Yao, F. Zhang, and N. Wang,
Nat. Commun. {\bf 7}, 12955 (2016).
%
\bibitem{Hosur2012}
P. Hosur, S. A. Parameswaran, and A. Vishwanath,
Phys. Rev. Lett. {\bf 108}, 046602 (2012).
%
\bibitem{Degiorgi1999}
L. Degiorgi,
Rev. Mod. Phys. {\bf 71}, 687 (1999).
\bibitem{Kimura2006}
S. Kimura, J. Sichelschmidt, J. Ferstl, C. Krellner, C. Geibel, and F. Steglich,
Phys. Rev. B {\bf 74}, 132408 (2006).
%
\bibitem{Carbotte2016}
J. P. Carbotte,
Phys. Rev. B {\bf 94}, 165111 (2016).
%
\bibitem{Shao2019}
Y. Shao, Z. Sun, Y. Wang, C. Xu, R. Sankar, A. J. Breindel, C. Cao, M. M. Fogler, A. J. Millis, F. Chou, Z. Li, T. Timusk, M. B. Maple, and D. N. Basov,
PNAS {\bf 116}, 1168 (2019).
%
\end{thebibliography}
\end{document}